\documentclass[aps,prl,twocolumn,superscriptaddress]{revtex4-1}
\usepackage{graphicx}

\begin{document}


\title{Internal pressure in superconducting Cu intercalated Bi$_2$Se$_3$}

\author{Amit Ribak}
\affiliation{Physics Department, Technion - Israel Institute of Technology, Haifa 3200003 , Israel}
\author{Khanan B. Chashka}
\affiliation{Physics Department, Technion - Israel Institute of Technology, Haifa 3200003 , Israel}
\author{Elias Lahoud}
\affiliation{Physics Department, Technion - Israel Institute of Technology, Haifa 3200003 , Israel}
\author{Muntaser Naamneh}
\affiliation{Physics Department, Technion - Israel Institute of Technology, Haifa 3200003 , Israel}
\author{Shahar Rinott}
\affiliation{Physics Department, Technion - Israel Institute of Technology, Haifa 3200003 , Israel}
\author{Yair Ein-Eli}
\affiliation{Department of Materials Science and Engineering, Technion - Israel Institute of Technology, Haifa 3200003 , Israel}
\author{Nicholas C. Plumb}
\affiliation{Swiss Light Source, Paul Scherrer Institut, CH-5232 Villigen, Switzerland}
\author{Ming Shi}
\affiliation{Swiss Light Source, Paul Scherrer Institut, CH-5232 Villigen, Switzerland}
\author{Emile Rienks}
\affiliation{Helmholtz-Zentrum Berlin, BESSY, D-12489 Berlin, Germany}
\author{Amit Kanigel}
\affiliation{Physics Department, Technion - Israel Institute of Technology, Haifa 3200003 , Israel}


\date{\today}

\begin{abstract}
Angle-resolved photoemission spectroscopy is used to study the band-structure of superconducting electrochemically intercalated Cu$_x$Bi$_2$Se$_3$. We find that in these samples the band-gap at the $\Gamma$ point is much larger than in pristine Bi$_2$Se$_3$. Comparison to the results of band-structure calculations indicates that the origin of this large gap is internal stress caused by disorder created by the Cu intercalation. We suggest that the internal pressure may be necessary for superconductivity in Cu$_x$Bi$_2$Se$_3$.  
\end{abstract}


\maketitle

Recently, superconductivity has been found in Cu intercalated Bi$_2$Se$_3$ \cite{Hor_Cava_PRL,Ando-PRL,Ando-PRB}. Since Bi$_2$Se$_3$ is a well studied 3D topological insulator (TI) \cite{Zhang-Bi2Se3_single_Dirac_cone,Xia-Bi2Se3}, it has been suggested that  Cu$_x$Bi$_2$Se$_3$ is a 3D time-reversal invariant topological superconductor (TSC) - a superconductor with full bulk gap and gapless surface Andreev bound states \cite{Qi_TSC2009,Qi_TSC2010}. The discovery motivated Fu and Berg to provide a sufficient criterion for identifying a 3D time-reversal-invariant TSC; a superconductor that has an odd-parity order parameter and that its Fermi-surface (FS) encloses an odd number of Time Reversal Invariant Momenta (TRIM) points is a 3D TSC \cite{Fu-Berg_TSC}. 

Although the discovery has evoked extensive research, the nature of superconductivity in this compound is not yet understood. Indications of non-trivial superconductivity were provided by point-contact spectroscopy, where the spectra were found to have zero bias conductance peaks (ZBCP) \cite{ZBCP-Ando,ZBCP-Kanigel}, interpreted as a signature of a dispersive Majorana mode \cite{Hsieh&Fu}. On the other hand, we have shown recently that Cu$_x$Bi$_2$Se$_3$ does not satisfy the Fu and Berg criterion since the superconducting samples always have an open FS which encloses two TRIM points \cite{Lahoud-OpenFS}. Furthermore, scanning tunnelling microscopy measurements revealed a U-shaped fully gapped spectrum without a ZBCP \cite{Niv-Levy-STM-S-wave}. 

In this paper we use angle-resolved photoemission spectroscopy (ARPES) to study the electronic band structure of superconducting Cu$_x$Bi$_2$Se$_3$ intercalated electrochemically (EC). We observe a major change in the band structure that may be a result of internal pressure. We support our findings by performing \textit{ab initio} band structure calculations. 

High quality single crystals of Bi$_2$Se$_3$ were prepared using the modified Bridgman method \cite{ZBCP-Kanigel}. The as-grown crystals have a carrier density of $n\simeq10^{17}cm^{-3}$. These carriers are believed to be the result of Se vacancies which are always present in the material \cite{Hor-p-typeBiSe}. Cu was then electrochemically intercalated into the pristine crystals. For the intercalation we used a 2-electrode cell configuration with Cu wire serving both as reference and counter electrode, while the Bi$_2$Se$_3$ crystal served as working electrode \cite{Ando-PRL}. The electrolyte used was 0.1M CuI (99.999\%, anhydrous) dissolved in CH$_3$CN (99.9\%, HPLC, dried). A constant current was applied for a suitable time to intercalate the desired amount of Cu. 

The samples become superconducting only after an appropriate annealing procedure. The samples were annealed in a vacuum sealed ampoule for a few hours and quenched. We found that annealing below 565$^\circ$c doesn't yield superconducting samples, and annealing at 595$^\circ$c yields samples with the highest superconducting shielding volume fraction.

The magnetization data were measured with a commercial SQUID (Cryogenic S700X).
The ARPES data were measured at the PGM beam line at the Synchrotron Radiation Center (SRC) (Stoughton, WI), the HRPES-SIS beamline at SLS, PSI (Villigen, Switzerland) and at the $1^3$ beamline at BESSY II (Berlin, Germany). All the samples were cleaved at low temperature (20K at SRC and PSI, ~1K at BESSY) in a vacuum better than 5$\times$10$^{-11}$ torr and measured at the same temperature. Each sample was measured for no longer than 6 h; within this time we did not observe any change in the chemical potential. The crystal structure of the single crystals  was characterized by x-ray diffraction (XRD) using a Bruker D8 diffractometer with Cu \textit{K$\alpha$} radiation, and FEI Tecnai T20 transmission electron microscope (TEM).

\begin{figure}[h!]
	\centering	
	\includegraphics[width=86mm]{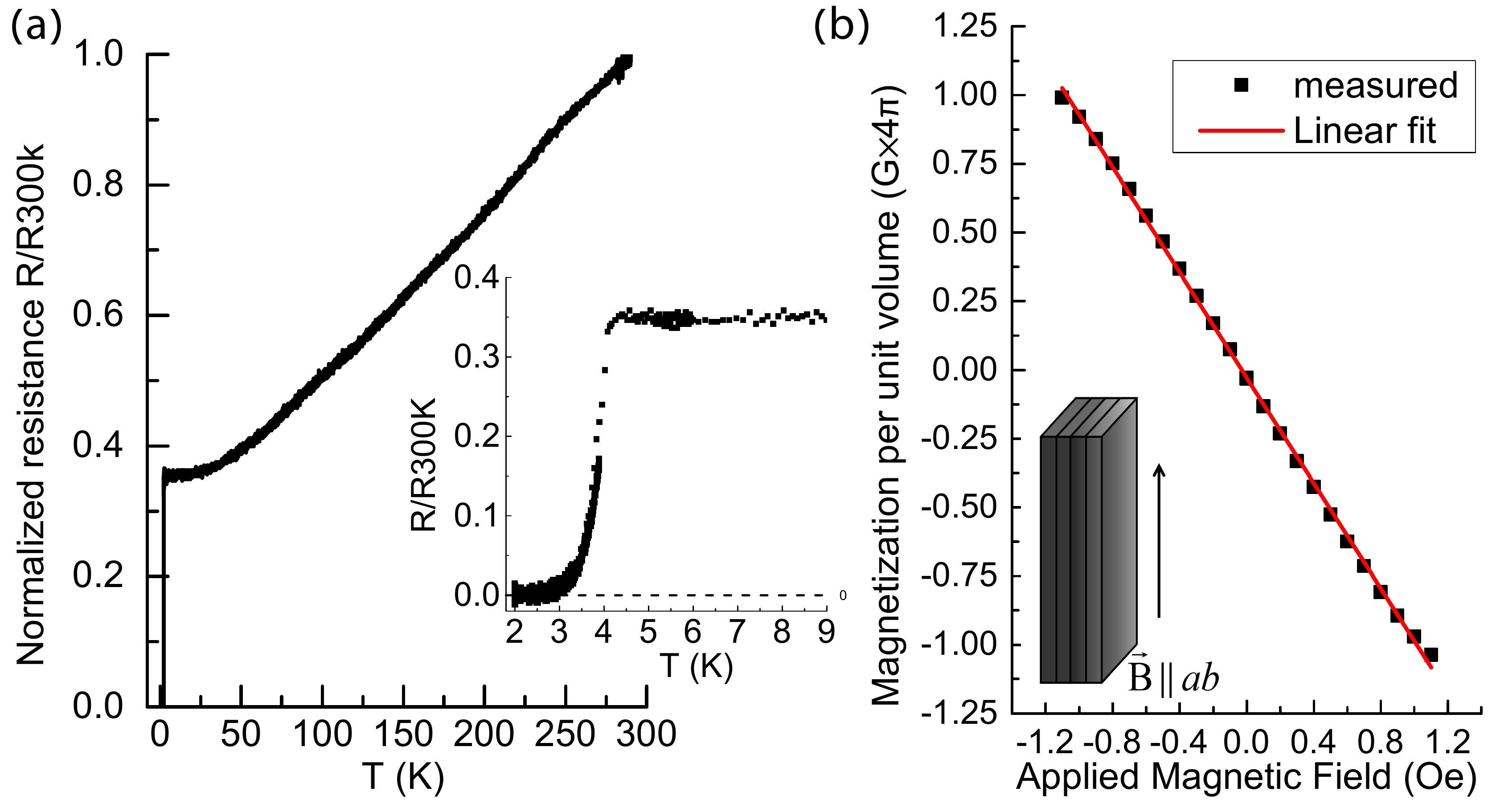}
	\caption{Transport and SQUID measurements. (a) Resistance as a function of temperature of Cu$_{0.27}$Bi$_2$Se$_3$ showing T$_{\textrm{c}}\approx$3.8K,  the resistance drops to zero below T$_c$. The Inset is a closeup on the superconducting transition. (b) Magnetization as a function of the applied magnetic field measured at 2K in ZFC. The linear behaviour persists up to $\sim$1.2G, and the slope represents  a ZFC shielding fraction of 90\%. }
	\label{RTSQUID}
\end{figure}

In Fig. \ref{RTSQUID}(a) we present a measurement of the resistance as a function of temperature of a typical EC intercalated sample with T$_{c}\approx$3.8K. 
For each sample, the magnetization as a function of the applied magnetic field was measured in the SQUID in zero-field cooling (ZFC) conditions at 2K. The samples were cut into rectangular shape and the magnetic field was applied parallel to the ab plane. From the linear part of the magnetization curve we extract the magnetic susceptibility and the ZFC shielding fraction. In most samples the linear part persists up to an applied field of $~$1.2G. In Fig. \ref{RTSQUID}(b) we show such a measurement with 90\% shielding fraction in ZFC.  

The EC intercalation of Cu into Bi$_2$Se$_3$ dopes the crystal with electrons. For superconducting (SC) samples with Cu concentration of about 25\% the carrier density is about $n\simeq10^{21}cm^{-3}$ as indicated by Hall measurements. The resulting ARPES spectra shows a rigid shift of the Dirac point to lower energies, while the surface state dispersion remains intact. The conduction band, in turn, becomes more populated with electrons. 

\begin{figure}[h!]
	\centering	
	\includegraphics[width=86mm]{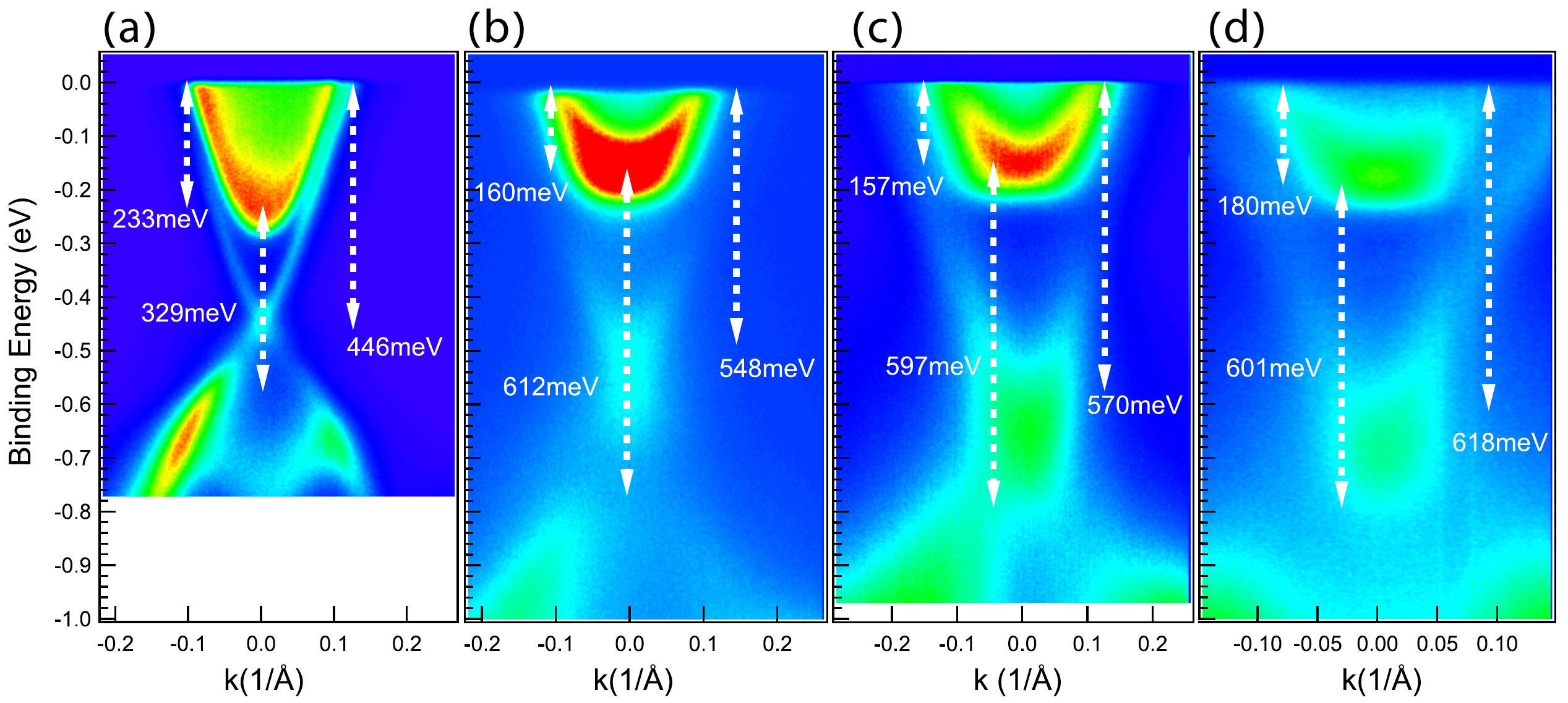}
	\caption{ARPES spectra showing the usual and the large gaps. The large gap spectra was seen in data obtained at different beam-lines using different photons energies. (a) Normal gap, PSI SLS h$\nu$=19eV, T=25K. (b) Large gap, PSI SLS, h$\nu$=20eV, T=25K. (c) Large gap, BESSY, h$\nu$=19eV, T=1K. (d) Large gap, SRC, h$\nu$=16eV, T=29K. }
	\label{NormalvsDouble}
\end{figure}

In most cases the ARPES spectra shows the usual bulk band gap of $\sim$300meV (Fig. \ref{NormalvsDouble}(a)).
In about 10\% of the spectra we observe a major difference in the band structure. In these spectra we see an increase in the bulk band gap to $\sim$600meV, about twice the gap of Bi$_2$Se$_3$. In Fig. \ref{NormalvsDouble} we compare the two kinds of spectra. As the bulk gap increases, the Dirac point shifts towards lower energies, and the bulk conduction band becomes more flat (Figs. \ref{NormalvsDouble}(b-d)).      
    
\begin{figure}[h!]
	\centering	
	\includegraphics[width=86mm]{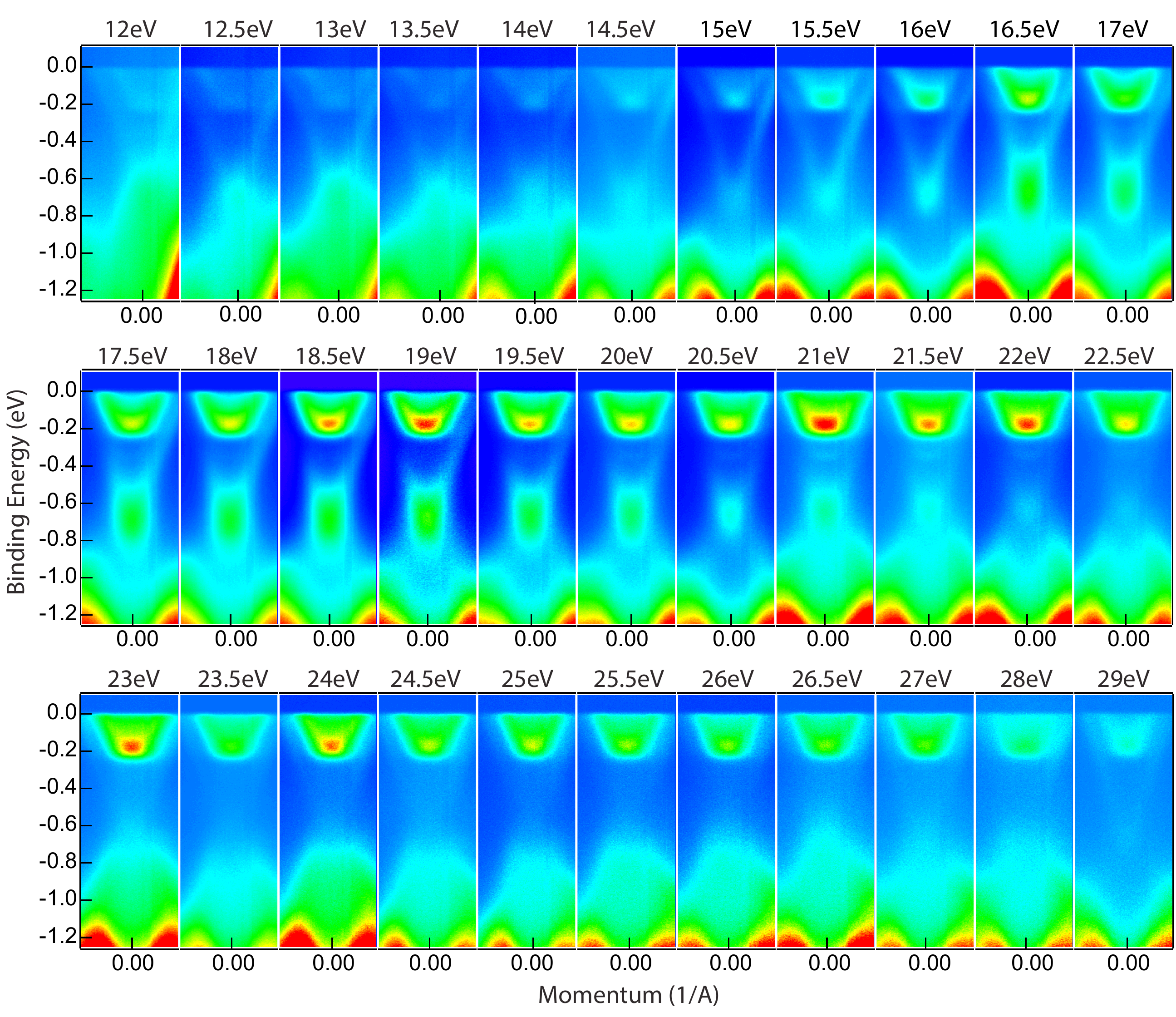}
	\caption{Photon energy dependence of the ARPES data for a large gap spectrum. All the scans were taken at 29K. We find that the bulk band is visible at the entire photon-energy range, which covers a momentum range larger than the $\Gamma$-Z separation.}
	\label{DGkz}
\end{figure}

In order to map the dispersion along the \textit{k$_Z$} direction in a case where a large gap spectrum was found, we performed ARPES measurements at normal emission over a wide range of photon energies in steps of 0.5 eV. In Fig. \ref{DGkz} we show a set of scans with different photon energies. For all the photon-energies the bulk conduction band is visible, and \textit{k$_F$} barely changes, indicating a weak dispersion along the $\Gamma$-$Z$ direction.   

\begin{figure}[h!]
	\centering	
	\includegraphics[width=86mm]{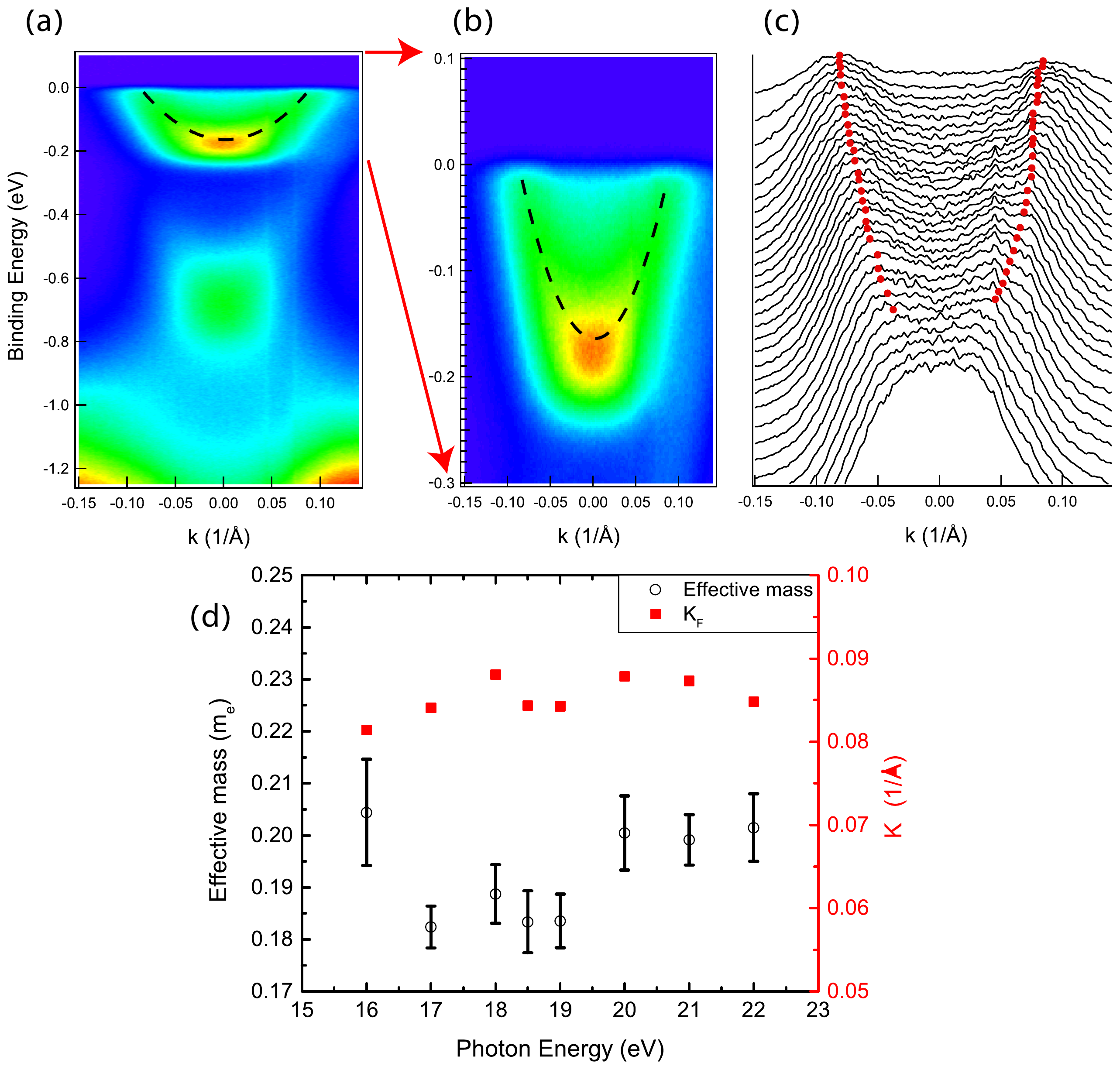}
	\caption{Effective mass of the bulk band. (a) Dispersion of a spectrum with a large gap taken at 19eV and a closeup on the conduction band (b). (c) MDCs for the 19eV photon-energy data. The red points are the maxima of the MDCs. We use these maxima to extract the dispersion. (d) Summary of the fit results. The red points (black circles) represent \textit{k$_F$} (effective mass) as a function of the photon energy.}
	\label{DGmass}
\end{figure}

Next, we look in greater detail on the band structure of a large gap spectrum. In these cases the shape of the conduction band is very clear; this can be seen in Figs. \ref{DGmass}(a) and \ref{DGmass}(b). By following the peaks in the momentum distribution curves (MDCs), we extract the band dispersion (Fig. \ref{DGmass}(c)). Fitting the data to a simple parabolic dispersion model, we can find \textit{k$_F$} and the effective mass at different photon energies. The best fit is shown as black dashed lines in Figs. \ref{DGmass}(a) and \ref{DGmass}(b). The effective masses resulting from these fits versus the photon energy are shown in Fig. \ref{DGmass}(d) together with the \textit{k$_F$}'s. We find that when moving away from the $\Gamma$-point towards the zone-boundary \textit{k$_F$} barely change, indicating that the dispersion in the \textit{k$_Z$} direction is very weak.

The EC intercalation process takes place at room temperature, thus the Cu atoms must intercalate into the van-der-Waals gaps between the Bi$_2$Se$_3$ layers. The intercalation leads to an inhomogeneous Cu distribution in the van-der-Waals gaps which reduces the uniformity of the crystal and creates disorder. This disorder results in internal stress which may induce the gap enlargement.
    
\begin{figure}[h!]
	\centering	
	\includegraphics[width=86mm]{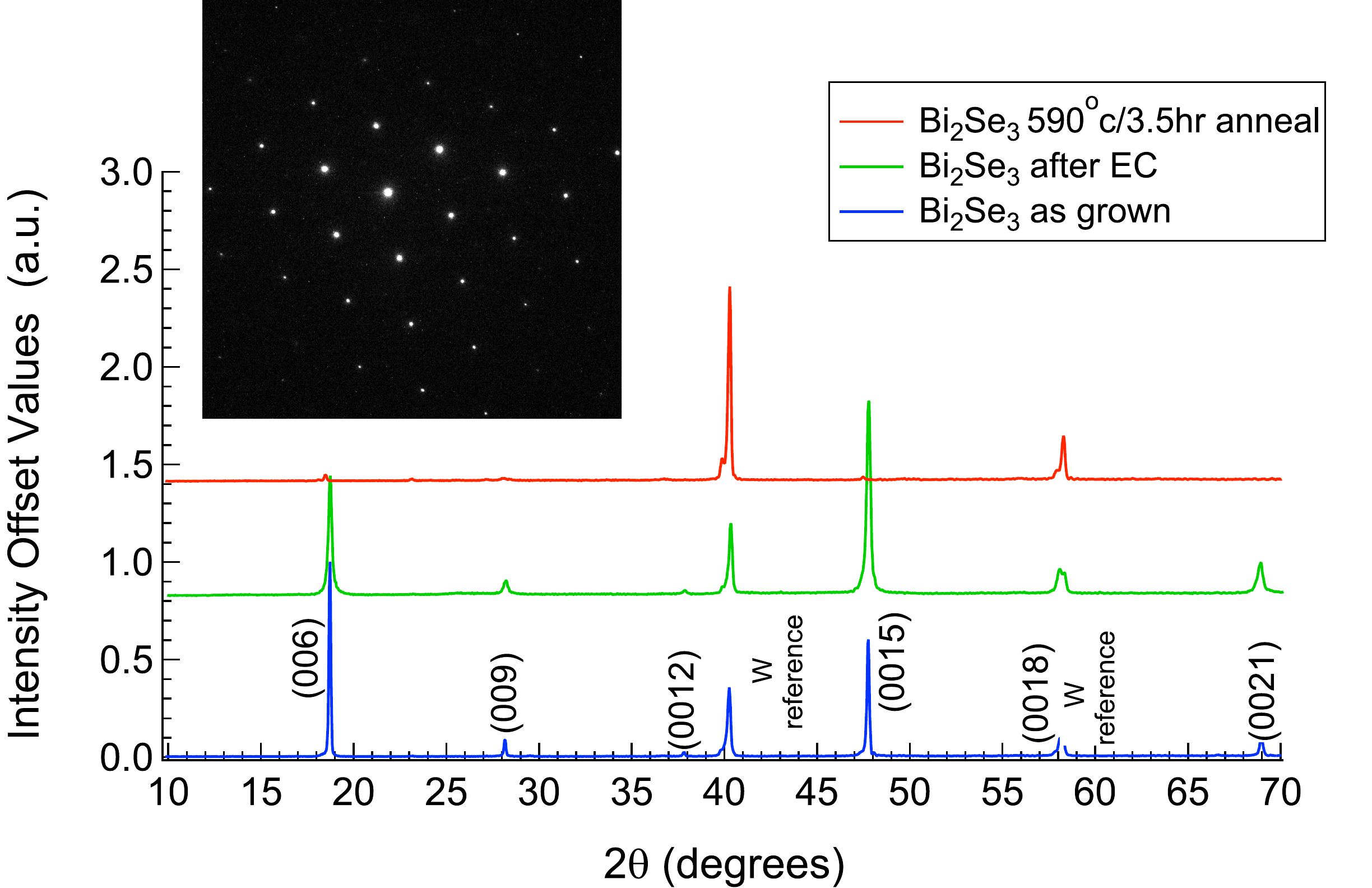}
	\caption{XRD spectra of a single crystal of superconducting Cu$_{0.23}$Bi$_2$Se$_3$. The blue curve shows peaks from as-grown Bi$_2$Se$_3$ on top of Tungsten (W) reference powder. The crystal is orientated such that only planes belong to the $(00l)$ family are visible. The electrochemical process (green curve) causes some broadening of the spectrum. After the annealing (red curve) the peaks are barely visible as the d-space along the c-axis varies significantly due to internal pressure. The inset presents the $[001]$ zone axis TEM diffraction of a double gap sample shown in Fig. \ref{NormalvsDouble}(b). The diffraction fits pristine Bi$_2$Se$_3$, thus the internal pressure is likely to be uniaxial along the c-axis.}
	\label{XRD}
\end{figure}

Evidence for this disorder can be seen in the XRD spectra in Fig. \ref{XRD}. These spectra were taken from a single crystal which was oriented such that only peaks belonging to the $(00l)$ family are visible. The crystal was measured at three stages: the as-grown Bi$_2$Se$_3$ (blue); after the annealing process (green) and after the annealing process (red) where the sample becomes SC. Careful examination of the spectra reveals two phenomena. First, the peaks become broader after the EC intercalation and annealing compared to the as-grown crystal, indicating that the d-spacing along the c-axis varies due to disorder. Second, there is a systematic shift of the peaks in the annealed sample to lower $\theta$ values, indicating an increase in the d-spacing in agreement with \cite{Hor_Cava_PRL}. 

In the inset we show a $[001]$ zone axis TEM diffraction pattern taken from fragments of the same sample shown in Fig. \ref{NormalvsDouble}(b). Electron dispersive spectroscopy confirms that there is Cu in the area where the diffraction was taken from. The diffraction shows the hexagonal structure of Bi$_2$Se$_3$, indicating that the in-plane structure remains unaffected. This leads us to the conclusion that the inhomogeneous Cu distribution causes uniaxial internal stress along the c-axis direction, while the in-plane structure remains unaffected.    

In order to understand the expected effect of such a stress distribution on the band structure we performed \textit{ab initio} calculations of the electronic structure of Bi$_2$Se$_3$ under uniaxial pressure along the c-axis. The calculations were carried out in the framework of the Perdew-Burke-Ernzerhof 
generalized gradient approximation of the density functional theory \cite{PBE}, as implemented in the Wien2K package \cite{Wien2K}.

\begin{figure}[h!]
	\centering	
	\includegraphics[width=86mm]{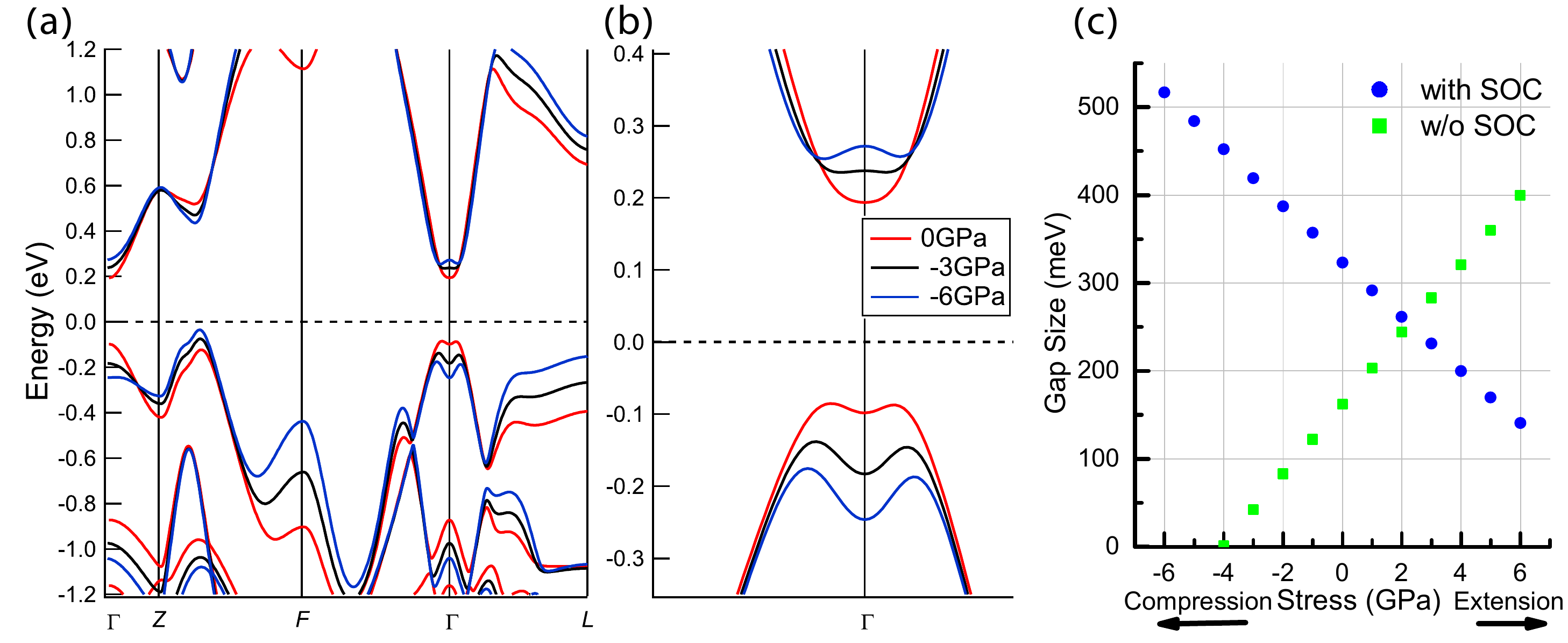}
	\caption{Band structure calculations. (a) The band structure at different \textit{k}-points. (b) Closeup on the band structure at the $\Gamma$-point. It can be seen that as pressure increases the bulk band gap at the $\Gamma$-point increases and the conduction band becomes more flat. (c) Calculated bulk band gap at the $\Gamma$-point versus stress. When SOC is taken into account (blue circles), the bulk band gap at the $\Gamma$-point increases with compression stress, and decreases with extension stress. Without SOC (green squares) the trend is opposite.}
	\label{DFT}
\end{figure}

Figs. \ref{DFT}(a) and (b) present the calculated band structure at ambient pressure (red) as well as under uniaxial compression pressure of 3 (black) and 6 (blue) GPa. When we apply uniaxial compression on the crystal the band gap increases and the conduction band becomes more flat. The results are very similar to the ARPES spectra presented in Figs. \ref{NormalvsDouble} and \ref{DGmass}.
The band gap at the $\Gamma$-point increases monotonically as we apply uniaxial compression pressure and decreases under extension. This behaviour is observed only when SOC is taken into account. As Fig. \ref{DFT}(c) shows, without SOC the trend is opposite.

As showed by Zhang \textit{et al.} \cite{Zhang-Bi2Se3_single_Dirac_cone}, the band gap size is determined by crystal-field splitting between different \textit{p} orbitals and by the spin orbit coupling (SOC) strength. A priori, the effect of uniaxial stress on the band structure could not be easily predicted. It might have happened that the crystal-field splitting would be affected more by the stress and the gap would decrease. Our calculation results in Fig. \ref{DFT}(c) show that when SOC is taken into account the gap increases, and the crystal remains in the non-trivial toplogical phase.     

Previous studies have shown that 3D topological insulators such as Bi$_2$Se$_3$ , Bi$_2$Te$_3$ and Sb$_2$Te$_3$ become superconducting when subjected to a large enough hydrostatic pressure \cite{Pressure_2011_PNAS,Pressure_2013_scin_rep,Pressure_2013_PRL,Pressure_2013_PRL,Pressure_2013_IOP}. Refs. \cite{Pressure_2013_PRL} and \cite{Pressure_2013_IOP} show that for Bi$_2$Se$_3$ the onset of superconductivity at pressure $\ge$11GPa is accompanied by a $\sim4$ orders of magnitude increase in the carrier density, much like the increase in the superconducting Cu$_x$Bi$_2$Se$_3$. 

The fact that Bi$_2$Se$_3$ becomes SC under pressure, the ARPES spectra and the band structure calculations, suggests that the role played by the Cu intercalation in inducting superconductivity in Bi$_2$Se$_3$ is not merely electron doping, but also creating stress along the c-axis. 

Shirasawa \textit{et al.} \cite{CuBi2Se3_films} studied the structure and transport properties of Cu-doped Bi$_2$Se$_3$ films. They report an increase in the carrier density and in the c-axis much like in the bulk crystals, but the films does not exhibit SC up to 13 quintuple layers. They conclude that in SC crystals the role played by the Cu is not only electron doping, and the inhomogeneity is important for SC.  

Our results are in agreement with a recent optical spectroscopy experiment which measured the change in the onset of interband transitions, $\omega_{mb}$, upon EC intercalation \cite{Burch_Optical_CuBiSe}. $\omega_{mb}$ measures the energy of the top of the valence band in respect to the Fermi-level. It was found that $\omega_{mb}$ is larger by  $\sim$400meV  in Cu$_{0.22}$Bi$_2$Se$_3$ compared to its value in the pristine samples. This increase is consistent with an increase of $\sim$100meV in the Fermi-energy due to doping and in addition a $\sim$300meV increase in the band-gap.   
This supports our conclusion that the bulk of the EC intercalated samples has a large gap and the fact that ARPES spectra showing this large gap are relatively rare is related  to stress relief on the surface. 

It is expected that cleaving the sample will relief the stress from the outer most layers which are probed by ARPES and other surface-sensitive techniques. This is probably the reason we
 find a large gap spectra in only $\sim10\%$ of the cleaves. 

Measurements such as point-contact spectroscopy and STM are performed on small, smooth flakes cleaved from a larger crystal.  These are exactly the cases where we expect the internal pressure to be relieved and this will lead to a non superconducting surface. At a lower temperature the surface can become SC through a proximity process.  
STM measurements at a temperature of 208mK found a fully gapped spectrum \cite{Niv-Levy-STM-S-wave}. 

In conclusion, we have shown that electrochemical intercalation can produce samples with almost full superconducting shielding fraction. ARPES data from EC intercalated Cu$_x$Bi$_2$Se$_3$ show a major change in the band structure compared to pristine Bi$_2$Se$_3$, in particular a large band gap is found at the $\Gamma$-point.  XRD and TEM measurements show that the in-plane structure remains intact while the crystals are disordered along the c-axis. Comparison of the ARPES spectra with band-structure calculations suggests that the large band gap is a result of internal pressure caused by inhomogeneous Cu distribution.
 We argue that superconductivity in Cu doped Bi$_2$Se$_3$ is not only a result of the electronic doping, but also a result of this internal stress.

\begin{acknowledgments}
We acknowledge the Paul Scherrer Institut, Villigen, Switzerland for provision of synchrotron radiation beamtime at beamline HRPES-SIS of the SLS.
We thank HZB for the allocation of synchrotron radiation beamtime.
The Synchrotron Radiation Center is supported by NSF DMR 0084402.
This research was supported by the Israel Science Foundation (grant No. 885/13) and the Swiss National Science Foundation (Grant No. 200021-137783).
The research leading to these results has received funding from the European Community's Seventh Framework Programme (FP7/2007-2013) under grant agreement n.$^{\circ}$312284.
We thank Ilia Khait for helping with the computational work. 
\end{acknowledgments}

\bibliography{Refrences}

\end{document}